\documentclass[11pt]{article}
\usepackage{theorem}
\usepackage{latexsym}
\usepackage{amsmath}

\usepackage[dvips]{graphicx}
\usepackage{graphicx}
\usepackage{subfigure}

\theoremheaderfont{\scshape}
\newtheorem{theorem}{Theorem}[section]
\newtheorem{corollary}{Corollary}[section]
\newtheorem{lemma}{Lemma}[section]
\newtheorem{proposition}{Proposition}[section]
\newtheorem{postulate}{Postulate}[section]
\newtheorem{statement}{Statement}[section]
{\theorembodyfont{\rmfamily}
\newtheorem{example}{Example}[section]}
{\theorembodyfont{\rmfamily}
\newtheorem{remark}{Remark}[section]}
{\theorembodyfont{\rmfamily}
\newtheorem{definition}{Definition}[section]}

\newcommand{\newc}{\newcommand}

\newc{\be}{\begin{equation}}
\newc{\ee}{\end{equation}}
\newc{\bea}{\begin{eqnarray}}
\newc{\eea}{\end{eqnarray}}
\newc{\beas}{\begin{eqnarray*}}
\newc{\eeas}{\end{eqnarray*}}

\newc{\pardt}{\partial_{t}}
\newc{\pardxi}{\partial_{i}}
\newc{\pardts}{\partial_{t^{*}}}
\newc{\pardxis}{\partial_{i^{*}}}
\newc{\pardxj}{\partial_{j}}
\newc{\pardxk}{\partial_{k}}
\newc{\pard}{\partial}
\newc{\ti}{\tilde}

\newc{\s }{\overline}
\newc{\tr }{\textrm}

\newc{\sect}{\section}
\newc{\subs}{\subsection}

\newc{\defi}{\definition}
\newc{\prop}{\proposition}
\newc{\rem}{\remark}
\newc{\lem}{\lemma}
\newc{\exa}{\example}
\newc{\theo}{\theorem}
\newc{\coro}{\corollary}
\newc{\post}{\postulate}
\newc{\state}{\statement}

\begin{document}
\baselineskip0.5cm
\renewcommand {\theequation}{\thesection.\arabic{equation}}
\title{Vortex dynamics in rotating counterflow and
plane Couette and Poiseuille turbulence in superfluid Helium}

\author{
D.~Jou$^1$, M. Sciacca$^2$ and
M.S.~Mongiov\`{\i}$^2$\thanks{Corresponding author.}}

\date{}
\maketitle
\begin{center} {\footnotesize
$^1$ Departament de F\'{\i}sica, Universitat Aut\`{o}noma de
Barcelona, 08193 Bellaterra, Catalonia, Spain \\
$^2$ Dipartimento di Metodi e Modelli Matematici Universit\`a di
Palermo, c/o Facolt\`{a} di Ingegneria, \\ Viale delle Scienze,
90128 Palermo, Italy}

PACS number(s): 67.25.dk, 47.37,+q
\end{center} \footnotetext{E-mail addresses: david.jou@uab.es (D. Jou), msciacca@unipa.it (M. Sciacca)  and mongiovi@unipa.it (M. S.
Mongiov\`{\i})}

\begin{abstract}
An equation previously proposed to describe the evolution of
vortex line density in rotating counterflow turbulent tangles in
superfluid helium is generalized to incorporate nonvanishing
barycentric velocity and velocity gradients. Our generalization is
compared with an analogous approach proposed by Lipniacki, and
with experimental results by Swanson {\it et al.} in rotating
counterflow, and it is used to evaluate the vortex density in
plane Couette and Poiseuille flows of superfluid helium.
\end{abstract}

\section{Introduction}
Many researches of quantum vortices in superfluids have been
carried out on rotating systems and counterflow situations, both
of them with vanishing barycentric velocity gradient
\cite{Do}--\cite{NF-RMP-67-1995}. Evolution equations have been
proposed to describe the influence of heat flux and of angular
velocity on the vortex dynamics \cite{JMPRB69-2004} generalizing
the well-known Vinen's equation for non-rotating systems
\cite{Do}--\cite{NF-RMP-67-1995, Vinen-PRSL-1957}. An interesting
challenge is to generalize these vortex evolution equations to
include the influence of barycentric flow, which has much
practical interest, for instance, in cryogenic applications. Here,
we carry out such a generalization and we examine a recent
proposal by Lipniacki \cite{LipniaEJMB25-2006}, which opens an
interesting perspective but which, on the other side, discloses
some aspects which have not been yet settled out with enough
clarity.

The aim of this paper is to generalize a previous equation
proposed for rotating counterflow superfluid turbulence
\cite{JMPRB69-2004} by emphasizing more explicitly the dynamical
role of the rotational of the superfluid velocity ${\bf v}_s$,
related to quantized vortices. This allows us to write a proposal
for the evolution equations of vortices in plane Couette and
Poiseuille flows. In Section~\ref{section3} we review some aspects
of rotating counterflow and compare our generalized expression
with Lipniacki's proposal \cite{LipniaEJMB25-2006}, which
underlines the role of the polarization rather than of
$\textrm{rot}\ {\bf v}_s$ itself, and we stress some open
problems. In Section~\ref{section4bis} we use a thermodynamic
formalism to relate the dynamical equation for vortices with a new
term appearing in the mutual friction force, which we use for a
comparison with that one by Lipniacki. In Section~\ref{section4}
we discuss several aspects of Couette and Poiseuille flows of
superfluid helium including the presence of quantized vortices.

\section{Rotational of superfluid velocity and the dynamics of vortex line density}
\setcounter{equation}{0} An evolution equation for the dynamics of
quantum vortices in rotating helium under counterflow was proposed
in \cite{JMPRB69-2004}, describing the influence of the heat flow
and of angular velocity on the vortex line density. In particular,
the vortex-line density $L$ was assumed to obey the
following equation 
 \be\label{deLsudt1}%
  {\tr d L \over \tr d t}= -\beta  \kappa  L^2 +
\left[\alpha_1 V+\beta_2 \sqrt{\kappa \Omega}\right]
L^{3/2}-\left[\beta_1 \Omega+\beta_4 V  \sqrt{ \frac{
\Omega}{\kappa}}\right] L, %
\ee
 where $\beta$, $\alpha_1$,
$\beta_2,$ $\beta_1,$ and  $\beta_4$ are dimensionless
coefficients, $\kappa=h/m$ is the quantum of vorticity ($m$ the
mass of the $^4$He atom and $h$ Planck's constant), $V=|{\bf V}|$
(with  ${\bf V}={\bf v}_n-{\bf v}_s$) is the counterflow velocity,
the relative velocity between averaged normal and superfluid
velocities, which is proportional to the heat flux across the
system, and $\Omega=|{\bf\Omega}|$ is the angular velocity of the
container. The values of the coefficients were obtained in
\cite{JMPRB69-2004} by comparison with experimental data of
\cite{SBD-PRL50} and they were seen to satisfy the relations
$\beta_4=\sqrt{2}\alpha_1$ and $\beta_1=\sqrt{2}\beta_2-2\beta,$
which are required on relatively general arguments about the form
of solutions. The values of the coefficients appearing in
(\ref{deLsudt1}) were independently calculated in \cite{ASC-2007},
and agree with those obtained in \cite{JMPRB69-2004}. When
$\Omega=0$, equation (\ref{deLsudt1}) reduces to the well-known
Vinen's equation \cite{Vinen-PRSL-1957}, with parameters
$\alpha_1$ and $\beta$ being respectively related to the
production and destruction of vortices per unit volume and time.

In \cite{JMPRB69-2004, JM-PRB72-2005} it was shown that the value
of coefficient $\alpha_1$ depends on the angle between the
counterflow velocity $\bf V$ and Schwarz's binormal vector ${\bf
I}$ \cite{SchwPRB38-1988} (see equation (\ref{I})). As observed in
\cite{JMPRB69-2004}, $\alpha_1=\alpha_V {\bf I\cdot \hat V}$, with
$\alpha_V$ the coefficient appearing in Vinen's equation (pure
counterflow) \cite{Do}.
 Schwarz derived Vinen's equation using the vortex filament
model obtaining $\alpha_V=\alpha c_1$, where $\alpha$ is the
well-known coefficient appearing in the expression of the mutual
friction force between vortex lines and the normal fluid and $c_1$
denotes the average curvature of the tangle (see equation
(\ref{c1})). In \cite{JMPRB69-2004} in the regime of high
rotation, the value ${\bf I\cdot \hat V}=1/2$ was found, so
indicating that the vortex tangle is highly polarized. Coefficient
$\beta$ is linked to the average squared curvature of the vortices
as $\beta\kappa=\alpha\tilde\beta c_2^2$, with $c_2^2$ defined in
equation (\ref{c1}) and $\tilde\beta$ the vortex tension
parameter, defined as $\epsilon_V=\kappa\rho_s\tilde\beta$ with
$\epsilon_V$ the energy per unit length of vortex line \cite{Do}.

These equations lack an important source of vorticity, namely a
barycentric velocity gradient, which is known to produce
turbulence in many actual flows. Thus, it would be useful to
generalize (\ref{deLsudt1}) by incorporating in it barycentric
velocity gradients. A possible way to do so would be simply adding
new terms basing on dimensional analysis and on comparison with
the observed phenomenology. Instead of proceeding in this way, we
will interpret (\ref{deLsudt1}) in some deeper terms, which will
be useful for a consistent incorporation of the velocity gradient.

To generalize equation (\ref{deLsudt1}) we  note that in the
particular case of pure rotation $\Omega$ is related to
$\textrm{rot}\ {\bf v}_s$ as $2\Omega = |\textrm{rot}\ {\bf
v}_s|$, where ${\bf v}_s$ is the macroscopic superfluid velocity.
As Lipniacki noted in a different proposal
\cite{LipniaEJMB25-2006}, writing an equation such as
(\ref{deLsudt1}) in terms of $\textrm{rot}\ {\bf v}_s$ and $V$
rather than in terms of $\Omega$ and $V$ would be more general,
because it would reduce to (\ref{deLsudt1}) for rotation, and it
could be applied to other flows as plane Couette or Poiseuille
flows (see Section \ref{section4}), where $|\textrm{rot}\ {\bf
v}_s| = d v_{sx}(z)/dz$, $x$ being the direction of the fluid
motion, $z$ the direction orthogonal to the parallel plates, and
$v_{sx}(z)$ the macroscopical superfluid velocity, depending only
on $z$.

In this way, the natural generalization of (\ref{deLsudt1}) would
be to rewrite it in terms of $\textrm{rot}\ {\bf v}_s$ as
\be\label{deLsudt-rot} {dL\over dt}= -\beta  \kappa  L^2 +
\left[\alpha_1 V+\frac{\beta_2}{\sqrt{2}} \sqrt{\kappa
|\textrm{rot}\ {\bf v}_s|}\right] L^{3/2}-\left[\frac{\beta_1}{2}
|\textrm{rot}\ {\bf v}_s|+\frac{\beta_4}{\sqrt{2}} V \sqrt{ \frac{
|\textrm{rot}\ {\bf v}_s|}{\kappa}}\right] L. \ee

Equation (\ref{deLsudt-rot}) reduces to (\ref{deLsudt1}) for pure
rotation. Besides that, expression (\ref{deLsudt-rot}) generalizes
(\ref{deLsudt1}) also on dynamical grounds. Note, indeed, that in
(\ref{deLsudt1}) it is
 assumed that $|\textrm{rot}\ {\bf v}_s|$ is equal to $2 \Omega$.
However, it will take some time for ${\bf v}_s$ to get these
values, by starting after some arbitrary initial state.  Thus,
whereas $\Omega$ in (\ref{deLsudt1}) is taken as an externally
fixed parameter, in (\ref{deLsudt-rot}) $\textrm{rot}\ {\bf v}_s$
is a dynamical quantity, which must be described by a suitable
evolution equation. Then, the form (\ref{deLsudt1})  will be
useful after some transient interval, whereas (\ref{deLsudt-rot})
is expected to be valid also for fast changes in ${\bf v}_s$.
Further, equation (\ref{deLsudt-rot}) can be applied also in
different situations, as plane Couette and Poiseuille flows. Thus,
equation (\ref{deLsudt-rot}) is the central point of this paper,
as it generalizes (\ref{deLsudt1}) both to a wider set of external
conditions and to a wider domain of dynamical variations.

Comparison with a similar approach by Lipniacki
\cite{LipniaEJMB25-2006} will be useful for a better understanding
of both approaches.  Lipniacki \cite{LipniaEJMB25-2006} has
essentially proposed to use as variable the  so-called "polarity
vector" (see also \cite{JM-PRB..} ), an important quantity in
vortex dynamics, which he linked to the rotational of the averaged
superfluid velocity
 \be\label{q=Lip}%
 {\bf p}={\bf <s'>}=\frac{\int {\bf s'}
d \xi}{\int d \xi}= \frac{\nabla \times {\bf v}_s}{\kappa L}. \ee

For pure rotation one has ${\bf p} ={\bf \hat\Omega}$ and $\nabla
\times {\bf v}_s = 2 {\bf \Omega}$; therefore, we may rewrite
equation (\ref{deLsudt1}) in terms of ${\bf p}$. Note that $|{\bf
p}| \in [0,1]$ measures the directional anisotropy of the tangent
to the vortex lines: in particular $|{\bf p}|=1$ for a system of
parallel vortices and $|{\bf p}|=0$ for isotropic tangles. Thus,
it is possible to express (\ref{deLsudt-rot}) in terms of $\bf p$
and to group the terms in it in a slightly different way, namely,
in two groups, one of them with the factor $VL^{3/2}$ and the
other one with $k L^2$, mimicking in some way the form of the
original Vinen's equation. In this way, we rewrite
(\ref{deLsudt-rot}) as

 \be\label{deLsudt2} {dL\over dt}= -\beta \kappa
L^2\left[1-\frac{\beta_2}{\beta} \sqrt{\frac{\Omega}{\kappa
L}}+\frac{\beta_1}{\beta} \frac{\Omega}{\kappa L}\right]+\alpha_1
VL^{3/2}\left[1-\frac{\beta_4}{\alpha_1}\sqrt{\frac{\Omega}{\kappa
L}}\right], \ee which, recalling $\beta_1=\sqrt{2}\beta_2-2\beta$
and the previously mentioned relation $2\Omega =|\textrm{rot}\
{\bf v}_s|$ implying $2\Omega/\kappa L=|{\bf p}|$,
(\ref{deLsudt2}) assumes the more compact form
 \be\label{deLsudtproposed} \frac{dL}{dt}=\alpha_1 V
L^{3/2}\left[1-A\sqrt{|{\bf p}|}\right]-\beta \kappa
L^2\left[1-\sqrt{|{\bf p}|}\right]\left[1-B\sqrt{|{\bf
p}|}\right],\ee
 where $B=\frac{\beta_1}{2\beta}$ and
$A=\frac{\beta_4}{\sqrt{2}\alpha_1}.$ In \cite{JMPRB69-2004},
coefficient  $B$ was found to be $0.89$ while coefficient $A$ is
not properly a constant but undergoes a small step from $1$ to
$1.004$ at the first counterflow critical velocity $V_{c1}$. In
this work, as already pointed out, we neglect this step assuming
$A=1$.

When inhomogeneities in the line density $L$ are taken into
account, the evolution equation for line density $L$ must include
a vortex density flux ${\bf J}^L$ \cite{MJ-PRB(2007)}
\be%
\frac{\pard L}{\pard t}+ \nabla\cdot {\bf J}^L
=\sigma^L,%
\ee where $\sigma_L$ stands for the production term given by  the
right-hand side of equation (\ref{deLsudtproposed}).  The form of
${\bf J}^L$ contains a convective contribution, $L {\bf v_L}$ with
${\bf v_L}$ the velocity of vortex lines with respect to the
laboratory frame, and a diffusive contribution. In some
situations, when the rate of variation of the perturbations is
higher than the reciprocal of the relaxation time of the diffusive
flux \cite{MJ-PRB(2007),JMS-Ph.Lett.A}, one must take ${\bf J}^L$
as an independent variable \cite{Vortex-flux}. Here, neglecting
the relaxation time of ${\bf J}^L$ and considering isothermal
situations, we take for ${\bf J}^L$  the following simple law,
where the diffusive contribution is analogous to Fick's diffusion
law
\be\label{J^L}%
{\bf J}^L=-\tilde D\nabla L+L {\bf v_L}.%
\ee The coefficient $\tilde D$ (of the order of $\kappa$
\cite{MJ-PRB(2007)},\cite{JMS-Ph.Lett.A}) is the diffusion
coefficient of vortex lines.

For a general hydrodynamic description, the evolution equations
for ${\bf v}_n$ and ${\bf v}_s$ are needed. In particular, the
evolution of ${\bf v}_s$ is necessary to describe the evolution of
$\textrm{rot}\ {\bf v}_s$ in equation (\ref{deLsudt-rot}). A set
of equations frequently used are the Hall-Vinen-Bekarevich-
Khalatnikov equations \cite{Do,Bek-Kal}, which in an inertial
frame are written as \bea\label{HVBK-sistema-vn} \rho_n
\frac{\pard {\bf v}_n}{\pard t}+\rho_n({\bf v}_n\cdot
 \nabla){\bf v}_n =
- {\rho_n\over\rho}\nabla p_n
 - \rho_s S  \nabla T + {\bf F}_{ns}+\eta\nabla^2 {\bf v}_n,    \\
 \rho_s  \frac{\pard {\bf v}_s}{\pard t} +\rho_s({\bf v}_s\cdot
 \nabla){\bf v}_s = - {\rho_s\over\rho}\nabla p_s +
\rho_s S  \nabla T - {\bf F}_{ns}  +\rho_s {\bf T}.
\label{HVBK-sistema-vs}\eea Here, $p_n$ and $p_s$ are effective
pressures, defined as $\nabla p_n=\nabla p+(\rho_s/2)\nabla V^2,$
$\nabla p_s=\nabla p-(\rho_n/2)\nabla V^2,$ $p$ the total
pressure, $S$ the entropy, $\eta$ the dynamic viscosity of the
normal component, and $\rho_s {\bf T}$ the vortex tension force,
which vanishes for rectilinear vortices and for isotropic vortex
tangles, but which may be relevant in other situations. In the
situations considered in this paper, we will have ${\bf T}={\bf
0}$.

To describe the motion we need an expression for ${\bf F}_{ns}$,
the mutual force between normal and superfluid components. The
usual expression by Hall, Vinen, Bekarevich and Khalatnikov is
\cite{Do}

\be\label{F_ns-HVBK}%
 {\bf F}_{ns}=\alpha \rho_s \kappa L\left[{\bf {\hat p}}\times[ {\bf p}\times
( {\bf V}-{\bf v_i})] + \frac{\alpha'}{\alpha}{\bf {\hat p}}\times
( {\bf V}-{\bf v_i})\right], %
\ee with $\alpha$ and $\alpha'$ being friction coefficient
depending on temperature, and ${\bf v_i}$ the "self-induced
velocity", which in the HVBK
 equations is approximated by
\be%
{\bf v_i}=\ti \beta \nabla \times {\bf {\hat p}}.%
\ee The expression for ${\bf F}_{ns}$ must be consistent with the
dynamics of $L$. In Section~\ref{section4bis} we will explore how
(\ref{F_ns-HVBK}) should be modified in order to be consistent
with the evolution equation (\ref{deLsudtproposed}), and in
Section~\ref{section4} we will combine the equations in an
analysis of plane Couette and Poiseuille flows in steady
conditions.

\section{Rotating counterflow}\label{section3}
\setcounter{equation}{0} In this Section, we investigate the
proposed equation (\ref{deLsudt-rot}) for a rotating superfluid
helium inside a cylindric container in the absence and in presence
of counterflow and we compare some results of our proposal with
those of Lipniacki \cite{LipniaEJMB25-2006}, and with the
experimental data of Swanson {\it et al.} \cite{SBD-PRL50}. These
authors considered a rotating container filled of helium II with
an external counterflow $\bf V$ parallel to the angular velocity
$\bf \Omega$ of the container. For high angular velocities, they
observed two critical counterflow velocities $V_{c1}$ and $V_{c}$
such that for $0\leq V \leq V_{c}$ the line density $L$ is
approximately independent of $V$, undergoing only a small step
(about $0.4 \%$) at the first critical velocity $V_{c1}$ whereas
for $V \geq V_{c}$ the line density $L$ grows with $V$. Here we
will neglect the small variation of $L$ at the first critical
velocity $V_{c1}$, because our proposal reduces to
(\ref{deLsudt1}) in this situation
--- which was already carried out in \cite{JMPRB69-2004}
--- and it is not necessary for the comparison with
Lipniacki's proposal because the latter is valid only for $V \geq
V_{c}$.

\subsection{Pure rotation}
First of all, we consider the simplest situation of a cylindric
container  rotating around its axis. It is known that when the
angular velocity $\Omega$ exceeds a critical value $\Omega_c$ and
the stationary state is reached, vortex lines parallel to the
rotation axis are present whose number density follows the law
$L=2\Omega/\kappa.$ The presence of these vortices may be
explained observing that when the container begins to rotate the
viscous normal fluid rotates with it, whereas the superfluid
remains initially at rest, due to its vanishing viscosity. In this
situation, the difference between ${\bf v}_n$ and ${\bf v}_s$ is
zero along the rotation axis, but it is maximum near the walls of
the container, that is, the counterflow velocity increases for
increasing distance from the axis. In this way the remnant
vortices, which are formed during the cooling of helium and which
are pinned to the walls, are influenced by the counterflow
velocity. This implies the growth of these vortices in agreement
with the dynamical description proposed by Schwarz. According to
this idea, vortices will grow near the walls, due to the relative
velocity between normal and superfluid velocity, and will migrate
towards the bulk of the system, forming in the stationary
situation a regular array of vortices parallel to the rotation
axis.

The presence of vortices couples the normal fluid and the
superfluid through the mutual friction force so that vortices are
dragged by the normal fluid, and the average superfluid velocity
${\bf v}_s$ becomes different from zero. This fact justifies the
relation $\nabla \times {\bf v}_s^\Omega=2 {\bf \Omega}$ and the
substitution of $2\Omega/\kappa L=|{\bf p}|$ in equation
(\ref{deLsudt1}). At the light of the new arguments, in the case
of pure rotation the vortex line density becomes
$L=2\Omega/\kappa$, which implies $|{\bf p}| \equiv 1$.

Consider now equation (\ref{deLsudtproposed}) in the case of pure
rotation, when the stationary solution is reached, that is
$V=<|{\bf v}_n-{\bf v}_s|> \approx 0$. In this case
(\ref{deLsudtproposed}) has two stationary solutions, $|{\bf
p}|=1$ and $|{\bf p}|=1/B^2.$ As one can easily verify, the
solution $|{\bf p}|=1$ is stable if $B<1$ and this is the case
because the coefficient $B$ was found to be $0.89$
\cite{JMPRB69-2004}. To describe the non-stationary regime, one
needs to introduce equations (\ref{HVBK-sistema-vn}) and
(\ref{HVBK-sistema-vs}) for the averaged normal and superfluid
velocities.

\subsection{Fast rotation and external counterflow}
Equation (\ref{deLsudtproposed}) can be also written as

\be\label{deLsudtproposed2} \frac{\tr d L}{\tr d
t}=L^{3/2}\left(1-\sqrt{|{\bf p}|}\right)
\left[\alpha_1V-\beta\kappa L^{1/2}\left(1-B\sqrt{|{\bf
p}|}\right)\right].\ee  As we have pointed out above,  pure
rotation is well described by (\ref{deLsudtproposed2}), because in
this situation ${\bf v}_s\equiv{\bf v}_s^\Omega$ and $|{\bf p}|=1$
is a stationary solution of (\ref{deLsudtproposed2}), meaning
complete polarization. The non-zero stationary solutions of
(\ref{deLsudtproposed2}) are \be\label{soluzi_q_L} |{\bf p}|=1
\qquad \textrm{and}  \qquad L^{1/2}=\frac{\alpha_1}{\beta
\kappa}V+B\sqrt{\frac{|\nabla \times {\bf v}_s|}{\kappa}}.  \ee To
study the stability of the solution $|{\bf p}| = 1$, we linearize
Eq.~(\ref{deLsudtproposed2}) for the perturbations.  In the
hypothesis that the perturbation $\delta$ does not modify the
vorticity $\vec{\omega}=\textrm{rot}\ {\bf v}_s$, the relation
$\delta |{\bf p}|=-(|{\bf p}|/L) \delta L$ is obtained, which
allows us to obtain the following evolution equation for the
perturbation $\delta L$

\be\label{dedelsudt} \left(\frac{\pard \delta L}{\pard
t}\right)_{|{\bf p}|=1}=\left[\frac{\alpha_1
V}{2L^{1/2}}-\frac{1}{2}\beta\kappa(1-B)L\right]\delta L.\ee From
the previous equation it follows that the solution $|{\bf p}| = 1$
is stable for $V$ less than

\be\label{Vcri} V_c=\frac{\beta}{\alpha_1}(1-B)\sqrt{|\nabla
\times {\bf v}_s|\kappa}, \ee which corresponds to the critical
velocity $V_{c}$ in the experiments of Swanson {et al.}
\cite{SBD-PRL50}. Note that if $B = 1$ in Eq.~(\ref{Vcri}), the
critical counterflow velocity for which the straight vortex lines
parallel to the rotation axis become unstable is zero. From an
experimental point of view this is not the case because a
nonvanishing critical velocity is observed, confirming the value,
$B = 0.89<1$, obtained in reference \cite{JMPRB69-2004}.

For counterflow velocity higher than the critical velocity
(\ref{Vcri}), the solution $|{\bf p}|=1$ becomes unstable, and the
line density $L$ assumes the value (\ref{soluzi_q_L}b) which
depends on $V$ and $|\textrm{rot}\ {\bf v}_s|$.

Now, we consider the second term in the right hand side of
(\ref{soluzi_q_L}b), namely $B\sqrt{\frac{|\nabla \times {\bf
v}_s|}{\kappa}}$. For low values of the counterflow velocity, the
vorticity is essentially due to the rotation, and therefore we put
$|\nabla \times {\bf v}_s|=2\Omega,$ recovering the results
obtained in \cite{JMPRB69-2004}.

\subsection{Comparison with Lipniacki's proposal}\label{section5}
Recently a hydrodynamical model of superfluid turbulence was
proposed by Lipniacki \cite{LipniaEJMB25-2006}, mainly with the
aim to studying the hydrodynamics of partially polarized tangles
arising in rotating counterflow or in plane Couette flow. Thus, it
is interesting to compare with his work, whose aims are similar to
ours.

Lipniacki writes Vinen's equation as
\be\label{deLsudtLipni2}%
\frac{d L}{d t}=\alpha L^{3/2}c_1(|{\bf p}|) {\bf I}\cdot {\bf
V}-\beta\alpha_2c_2^2(|{\bf p}|)L^2,%
\ee where $\beta$ is a constant of the order of $\kappa$, and
$\alpha$ the friction coefficient appearing in the expression of
the mutual friction force; ${\bf I}$ is the binormal vector,
\be\label{I}%
{\bf I}   =\frac{<{\bf s'}\times {\bf s''}>}{<|{\bf
s''}|>},%
\ee defined by Schwarz  \cite{SchwPRB38-1988} to describe the
polarization of the binormal ${\bf s'}\times{\bf s''},$ of the
vortex lines, with ${\bf s'}$ and ${\bf s''}$ being the first and
second derivatives of the curve ${\bf s}(\xi)$ describing a vortex
line with respect to the arc-length $\xi$, ${\bf s'}$ the unit
tangent along the line and ${\bf s''}$ the curvature vector.

The coefficients $c_1$ and $c_2^2$ measure the average curvature
and curvature squared of the tangle, respectively. They are given,
according to the microscopic model by Schwarz
\cite{SchwPRB38-1988}, by
\be\label{c1}%
c_1=\frac{1}{\Lambda L^{3/2}}\int |s''| d\xi, \qquad
c_2^2=\frac{1}{\Lambda L^2}\int
|s''|^2 d\xi,%
\ee where $\Lambda$ is the volume on which one makes the averaging
indicated in (\ref{c1}). Lipniacki proposes that $c_1$ and $c_2^2$
should depend on the polarization $|{\bf p}|$, and that they
should vanish for completely polarized tangles because in this
case ${\bf s''}=0$ for all the vortex lines. To describe the
reduction in $c_1$ and $c_2^2$ with respect to its usual variable
for a nonpolarized tangle, which will be designed as $c_{10}$ and
$c_{20}^2 $, respectively, he assumes that
\be\label{c1Lip}%
c_1(|{\bf p}|)\simeq c_{10}\left[1-|{\bf p}|^2\right], \qquad
c_{2}^2(|{\bf p}|)\simeq c_{20}^2\left[1-|{\bf p}|^2\right]^2.%
\ee
 In contrast, our expression (\ref{deLsudtproposed}) could be
interpreted in this perspective as
\be\label{c1-nostro}%
 c_1(|{\bf
p}|)\simeq c_{10}\left[1-\sqrt{|{\bf p}|}\right], \qquad
c_2^2(|{\bf p}|)\simeq c_{20}^2\left[1-\sqrt{|{\bf
p}|}\right]\left[1-B\sqrt{|{\bf p}|}\right].%
\ee Therefore, it arises the question of the comparison of both
equations (\ref{deLsudtproposed}) and (\ref{deLsudtLipni2}) with
the experimental data, and a deeper understanding of the influence
of polarity on the coefficients $c_1$ and $c_2^2$.
\begin{figure}[h]
\includegraphics[width=12cm]{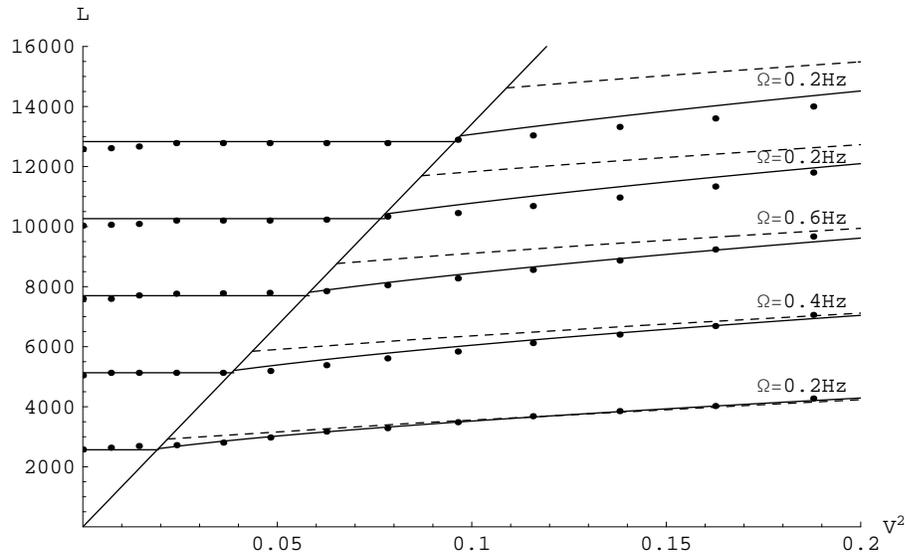}
\caption{{ Comparison of the stationary solutions of Lipniacki's
model (\ref{deLsudtLipni}) (dashed line) and Jou and
Mongiov\`{i}'s model (\ref{deLsudtproposed}) (black line) with the
experimental data (solid circles) by Swanson {\it et al.} for
counterflow velocity bigger than the second critical velocity
$V_{c}$ and angular velocity $0.2\ \textrm{Hz}, 0.4\ \textrm{Hz},
0.6\ \textrm{Hz}, 0.8\ \textrm{Hz} \ \textrm{and}\ 1\
\textrm{Hz}$. Lipniacki's model does not give the horizontal part
of the plot, corresponding to $V<V_{c}$.}}
  \label{confronto}
\end{figure}

The evolution equation for the vortex line density $L$, proposed
by Lipniacki in \cite{LipniaEJMB25-2006},  has the explicit form
\be\label{deLsudtLipni} %
\frac{dL}{dt}=\tilde \alpha I_0 c_{10} V L^{3/2}\left[1-|{\bf
p}|^2\right]-\tilde\beta \alpha c_{20}^2 L^2\left[1-|{\bf
p}|^2\right]^2,%
\ee where $I_0={\bf I\cdot \hat V }$, and the subscript $0$ stands
for independence of $I_0$ on $V$ and $L$. The author chooses for
$I_0$ the same values found in pure counterflow, in such a way to
not consider the anisotropy of the vortex tangle, which is present
owing of the high values of rotation considered in the experiments
by Swanson {\it et al.} \cite{SBD-PRL50}.

Equation (\ref{deLsudtLipni}), as the author remarks, does not
describe any of the two critical velocities, $V_{c1}$ or $V_{c}$,
of the experiments of Swanson {et al.} \cite{SBD-PRL50}.
 Lipniacki's aim is instead to describe the relation  between angular
velocity, counterflow, and line-length density for polarized
tangles above the second critical velocity $V_{c}$. This implies
the need of a comparison, in the uniform steady rotation and
counterflow, between (\ref{deLsudtproposed}), (\ref{deLsudtLipni})
and  the experimental data of Swanson {et al.} \cite{SBD-PRL50}.

The stationary solutions of the equation (\ref{deLsudtLipni}) are
$|{\bf p}|=1$  (which however is unstable) and
\be\label{L-tresol-Lip}
L=\frac{L_H}{\left(1-\left(L_\omega/L\right)^2\right)^2},%
\ee where
\be\label{Lh}%
L_H=V^2\left(\frac{c_{10}I_0}{\beta c_{20}^2}\right)^2 \quad
\textrm{and} \quad L_\omega=\frac{|\textrm{rot} {\bf
v}_s|}{\kappa}=\frac{2\Omega}{\kappa} %
\ee are the steady state vortex-line density in pure counterflow
and in pure rotation, respectively.

In Fig.~\ref{confronto}, we compare the results of equations
(\ref{deLsudtproposed}) and (\ref{deLsudtLipni}) with the
experimental data of the Fig.~2 of Swanson's experiments. It
follows that (\ref{deLsudtproposed}) (black line) describes better
the experimental data (solid circle) than (\ref{deLsudtLipni})
(dashed line), not only for $V>V_{c}$, but it also yields the
horizontal branch of the results for $V<V_{c},$ which are not
described by equation (\ref{deLsudtLipni}). Comparison with
experimental data shows that in the considered range of values of
$V$ and $\Omega$ equation (\ref{deLsudtproposed}) fits better the
experimental results.

A reason for the difference between proposals
(\ref{deLsudtproposed}) and (\ref{deLsudtLipni}) could be related
not to the evaluation of the integrals in (\ref{c1}) but to a
different microscopical interpretation of some terms in the
evolution equation for $L$. Schwarz's derivation
\cite{SchwPRB38-1988} is based on the dynamics of vortex breaking
and reconnection, and its production and destruction terms tend to
zero for completely polarized systems, as rightly pointed out by
Lipniacki. However, the origin of the rotational terms in
(\ref{deLsudt1}) could be completely different. It is known that
in rotating superfluid helium the vortices grow near the walls due
to the rotation, and drift towards the center of the system, where
they find a repulsion due to other vortices. These forces are
different from zero even for completely polarized vortices, in
contrast  to the terms from (\ref{c1}). It could then be that the
vanishing of the terms in (\ref{deLsudtproposed}) as
$1-\sqrt{|{\bf p}|}$ had a different physical origin than the
vanishing proposal by Lipniacki from a different model. These open
questions stress the need of the inclusion of rotational effects
in a more general version of Schwarz's derivation of Vinen's
equation.

\section{Thermodynamic analysis of polarized superfluid
 turbulence}\label{section4bis} \setcounter{equation}{0}
In this Section, we will perform two modifications of the
expression of the mutual friction force, as used in the HVBK
model, which are necessary to incorporate the anisotropy of the
vortex tangle and to insure the thermodynamic consistency of the
evolution equation for $L$ and for ${\bf v}_s$, according to the
formalism of linear irreversible thermodynamics
\cite{JM-PRB72-2005,Nemi_PRB154(1993)}. Since
(\ref{deLsudtproposed}) differs from the usual Vinen's equation,
it is logical to ask how these modifications will change the form
of ${\bf F}_{ns}$. For the sake of simplicity, we will neglect
here the contribution of the self-induced velocity in
(\ref{F_ns-HVBK}).

First, we will take into account the anisotropy of the tangle
introducing  the tensor ${\bf\Pi}={\bf\Pi}^s+{\bf\Pi}^a$, studied
in \cite{JM-PRB72-2005}, \cite{JM-PRB..},
 \be%
  {\bf\Pi}^s
\equiv{3\over 2} <{\bf U} -{\bf s'\bf s'}>, \hskip0.3in {\bf\Pi}^a
\equiv{3\over 2}{\alpha'\over\alpha} <{\bf W}\cdot {\bf s'}>.%
\ee
 In this equation $\bf s'$ is the unit vector tangent to the vortex
lines, $\bf s's'$ is the diadic product,  $\bf U$ is the unit
matrix, $\bf W$ is the Ricci third-order tensor and the angular
brackets stands for the average over vortex lines in a given
volume. The tensor ${\bf\Pi}^s$ describes the orientation of the
tangents ${\bf s'}$ of the vortex lines, and the tensor
${\bf\Pi}^a$ --- associated to an axial vector --- describes the
polarization; in other words, ${\bf\Pi}^a$ is related to the
first-order moment of the orientational distribution function of
${\bf s'}$ and ${\bf\Pi}^s$ is related to second-order moment. As
shown in Ref. \cite{JM-PRB..}, using tensor $\bf\Pi$, the
 mutual friction force can be written
\be\label{F_ns-isotr}%
{\bf F}_{ns} =- \alpha\rho_s\kappa
L{2\over3}{\bf \Pi}\cdot {\bf V}.%
\ee If we suppose isotropy in the tangle, it results ${\bf\Pi}^s =
{\bf U}$, ${\bf \Pi}^a = {\bf 0}$ and one finds the usual
expression
\be\label{devsdt2}%
{\bf F}_{ns}  =-{2\over3} \alpha \rho_s \kappa L {\bf V}.%
\ee

The tensor ${\bf \Pi}$ in (\ref{F_ns-isotr}) allows one to deal
under a same formalism an array of parallel straight vortices as
well as an isotropic tangle, and also the intermediate situations.

Now, we follow the general lines of \cite{JM-PRB72-2005},
\cite{Nem} to propose a modification to (\ref{F_ns-isotr}) with
the aim to determine an evolution equation for ${\bf v}_s$
consistent with (\ref{deLsudtproposed}).  According to the
formalism of nonequilibrium thermodynamics one may obtain
evolution equations for ${\bf v}_s$ and $L$ by writing ${d {\bf
v}_s/ dt}$ and ${dL/ dt}$  in terms  of  their conjugate
thermodynamic forces $-\rho_s {\bf V}$ and $\epsilon_V$. The
evolution equation (\ref{HVBK-sistema-vs}) for ${\bf v}_s$,
neglecting inhomogeneous contributions of pressure, temperature
and velocity, in an inertial frame, is written
\be\label{devsdt}%
\rho_s{d {\bf v}_s \over d t}  =-{\bf F}_{ns} =
  \alpha\rho_s\kappa L{2\over3}{\bf
\Pi}\cdot {\bf V}.%
\ee However, in the right-hand side of (\ref{F_ns-isotr}) must be
included additional contributions to make (\ref{devsdt})
thermodynamically consistent with (\ref{deLsudtproposed}).

In a way similar to that presented in \cite{JM-PRB72-2005}, we
write $d {\bf v}_s /d t$ and $d L/d t$ in matrix form in the
system (\ref{matrix}). In it, we write the equation for $L$ in the
form given in equation (\ref{deLsudtproposed}) and by means of
Onsager-Casimir reciprocity we obtain an additional contribution
to the evolution equation for ${\bf v}_s$. The result is
\be\label{matrix}%
\begin{pmatrix}
  {d {\bf v}_s\over dt}   \\
  {d L\over dt}
\end{pmatrix}
= L
\begin{pmatrix}
  - {1\over\rho_s} \alpha\kappa {2\over3}{\bf
\Pi} &
  \pm{\alpha_1\over\rho_s}L^{1/2}
 \left(1 -\sqrt{|{\bf p}|}
\right){\bf\widehat V} \\
-{\alpha_1 \over\rho_s}L^{1/2}
 \left(1 -\sqrt{|{\bf p}|}\right) {\bf\widehat V} & - {1 \over
 \epsilon_V}L
 \left(1 -\sqrt{|{\bf p}|}\right) \left(1 -B\sqrt{|{\bf p}|}\right)
\end{pmatrix}
\begin{pmatrix}
-\rho_s{\bf V} \\ \epsilon_V
\end{pmatrix}%
\ee The sign ambiguity present in that equation comes in a natural
way from the Onsager-Casimir reciprocity relation. Indeed, in
Feynman-Vinen view, $L$ is a scalar quantity which does not change
under time reversal, unlike the superfluid velocity ${\bf v}_s$
which changes sign. According to Onsager-Casimir, this leads to
antisymmetry of crossed coefficients thus leading to the + sign.
In Schwarz view, $L$ possesses vectorial properties and it would
change on time reversal, just like the superfluid velocity. This
leads to the symmetry of the kinetic coefficients in the matrix in
(\ref{matrix}), i.e. to the - sign in the upper right-hand term.
Below, we will directly take the minus sign, for the sake of a
more direct comparison with the work by Lipniacki.

Therefore the equation for $d{\bf v}_s/dt$ becomes
\be\label{devsdt4}%
\rho_s{d {\bf v}_s \over d t}  =\alpha\rho_s\kappa L{2\over3}{\bf
\Pi}\cdot {\bf V} -{\epsilon_V}
 {\alpha_1} L^{1/2}\left(1 -\sqrt{|{\bf p}|}\right){\bf\widehat
V}.%
\ee The new term not contained in  the evolution equation
(\ref{devsdt}) for ${\bf v}_s$ is the coupling term between $d{\bf
v}_s/dt$ and $\epsilon_V$ in the matrix in (\ref{matrix}). Note
that this term depends on the direction but not on the modulus of
$V$. This class of terms are called dry-friction terms.

Observing that in the steady state ($L$, $|\tr {rot} {\bf v}_s|$
and $\bf V$ constant) the solutions of vortex line density
equation (\ref{deLsudtproposed}) can be written as
\be\label{L_1/2a}%
L^{1/2}=\sqrt{|\tr {rot} {\bf v}_s|\over\kappa} , \hskip0.4in
\hbox{ for}~~
0<V<V_{c},%
\ee
\be\label{L_1/2}%
L^{1/2}= {\alpha_1\over\beta  \kappa}(V-V_{c}) + \sqrt{|\tr {rot}
{\bf v}_s|\over
\kappa},\hskip0.3in \hbox{for}~~V>V_{c},%
\ee and substituting them in (\ref{devsdt4}), we obtain the
following expression for the coupling force
\be\label{F_coupl_min}%
{\bf F}_{coupl}=-{\epsilon_V} \alpha_1 \left[ L^{1/2} -
{\sqrt{|\tr {rot} {\bf v}_s|\over \kappa}} \right]{\bf \widehat
V}= 0 , \hskip0.5in \hskip0.5in\hbox{
for}~~V<V_{c},%
\ee
\be%
{\bf F}_{coupl}=-{\epsilon_V} \alpha_1\left[ L^{1/2} - {\sqrt{|\tr
{rot} {\bf v}_s|\over \kappa}} \right]{\bf \widehat V} =
{\epsilon_V } {\alpha_1\over\beta  \kappa} (V-V_{c}) {\bf \widehat
V},
\hskip0.4in \hbox{ for}~~V>V_{c}. %
\ee As a consequence, for $V<V_{c}$ the coupling force is absent
(as in pure rotation) while, for $V>V_{c}$, when the array of
rectilinear vortex lines becomes a disordered tangle, the
additional term (\ref{F_coupl_min}) appears. Indeed, in a
almost-steady state ($L$ and $|\tr {rot} {\bf v}_s|$ constant),
for $V<V_{c}$, equation (\ref{devsdt}) would be valid, with $L$
expressed by (\ref{L_1/2a}), whereas, for $V>V_{c}$ it would
become
\be%
{d {\bf v}_s \over d t}  = \alpha\kappa L {2\over3} {\bf\Pi}\cdot
{\bf V} +{\epsilon_V }{\alpha_1 \over \beta \kappa \rho_s}
(V-V_{c}) {\bf \widehat
V},%
\ee with $L$ expressed by (\ref{L_1/2}). Summarizing, in steady
states for $V<V_{c}$ the dry-friction force is absent, while it
appears for $V>V_{c}$, when the array of rectilinear vortex lines
becomes a disordered tangle. Thus $V_{c}$ indicates the threshold
not only of the vortex line dynamics but also of the friction
acting on the velocity ${\bf v}_s$ itself; this seems logical, as
both variables are mutually related.

Summarizing, in this Section we have proposed to substitute the
expression (\ref{devsdt2}) of the mutual friction force used in
the HVBK model with
\be\label{F_ns-minus}%
{{\bf F}_{ns}}  =-\alpha\rho_s \kappa L{2\over3}{\bf \Pi}\cdot
{\bf V} -{\epsilon_V} {\alpha_1} L^{3/2}\left(1 -\sqrt{|{\bf
p}|}\right){\bf\widehat
 V}%
\ee for the sake of thermodynamic consistency with
(\ref{deLsudtproposed}).

To complete the comparison between Lipniacki's and our model, we
analyze in both models the expression of the mutual friction
force, which in HVBK equation is expressed by (\ref{F_ns-HVBK}),
while in general terms it is expressed as
\be\label{F_ns-in-gene}%
{\bf F}_{ns}=
\alpha\rho_s\kappa L<{\bf s'}\times[{\bf s'}\times ({\bf  V} -{\bf
v_i})]> + \alpha'\rho_s\kappa L <{\bf
s'}\times( {\bf V} -{\bf v_i})> . %
\ee Lipniacki neglects the coefficient $\alpha'$ in Eq.
(\ref{F_ns-in-gene}), and he approaches the quantity $<{\bf
s'}\times ({\bf s'}\times {\bf V})> \simeq [<{\bf s'}{\bf
s'}>-{\bf U}]{\bf V} ={\bf I}_v-{\bf V}$ (where $ {\bf I}_v=< {\bf
s'}({\bf s'}\cdot{\bf V} )>$) with
\be%
<{\bf s'}\times({\bf s'}\times {\bf  V})> \simeq {\bf
p}\times({\bf p}\times{\bf  V} )
 -  {2\over 3}(1-|{\bf p}|^2)  {\bf  V},%
\ee
 and the quantity
$<{\bf s'}\times({\bf s'}\times{\bf v_i})>
 \simeq
\tilde\beta <{\bf s'}\times{\bf s''}> =  \tilde\beta c_1
L^{1/2}{\bf I}$ with
\be%
<{\bf s'}\times({\bf s'}\times{\bf v_i})> \simeq
 -\tilde\beta I_0c_{10}(1-|{\bf p}|^2) L^{1/2}{\bf\hat
V}. %
\ee In explicit terms he uses
\be\label{F_ns-in-gene2}%
{\bf
F}_{ns}=\alpha\kappa\rho_s L\left[{\bf p}({\bf p}\cdot {\bf V}) -
{\bf V} { 2+|{\bf p}|^2\over 3} +\beta I_0 c_{10}(1-|{\bf
p}|^2) L^{1/2}{\bf \hat V} \right].%
\ee So in the work of Lipniacki,  the tensor ${2\over 3}{\bf
\Pi}^s=$ $<{\bf U}-{\bf s'} {\bf s'}>$ assumes the expression:
\be\label{Pis}%
{2\over 3}{\bf\Pi}^s\simeq [{\bf U}-{\bf p}{\bf p}] + {2\over 3}
(1-|{\bf p}|^2){\bf U}= {5-2|{\bf p}|^2\over 3}  {\bf
 U}-{\bf p}{\bf p}.%
\ee Note that (\ref{Pis})  does not respect the relation $trace [<
{\bf U}-{\bf s'}{\bf s'}>]=2$, following from the normalized
character of ${\bf s'}$, if $|{\bf p}|\ne 1$. In fact it is
\be%
\textrm{trace} \left[{5-2|{\bf p}|^2\over 3} {\bf U}-{\bf p}{\bf
p}\right] = 5-3|{\bf p}|^2. %
\ee The last term in (\ref{F_ns-in-gene2}) is a consequence of the
drift of the tangle in the direction of the counterflow, caused by
its anisotropy, where ${\bf I} = I_0 {\bf \hat V}$. This term is
substituted in our model by the last term in (\ref{devsdt4}),
which we can rewrite, recalling that
$\epsilon_V=\rho_s\kappa\tilde\beta$ and $\alpha_1={\alpha c_{10}}
I_0$ as
 \be%
 {\bf F}_{coupl} =-{\epsilon_V }
 {\alpha_1} L^{3/2}\left(1 -\sqrt{|{\bf p}|}\right){\bf\widehat V}
 = - {\rho_s\kappa\tilde\beta} {\alpha c_{10}} I_0 L^{3/2}
 \left(1 -\sqrt{|{\bf p}|}\right){\bf\widehat V}.%
\ee As it is seen, this term differs from the one of Lipniacki, in
the contribution due to the polarization of the tangle, which in
our approach depends on  $1-\sqrt{|{\bf p}|}$, and in Lipniacki's
one on $1-{|{\bf p}|^2}$. We note also that, in this
interpretation, we must choose the negative sign in the expression
of this coupling term, in agreement with the microscopic
derivation of the filament model by Schwarz.

Lipniacki does not consider the tension $\bf T$. For a comparison
with our work, we must observe that in Lipniacki's model the
quantity $<{\bf s's'}>$ is approximated by $\bf pp$, and this
approximation is correct only if most of the vortex lines in the
volume have the same direction.

In Ref. \cite{JM-PRB..} we have provided a microscopic
paramagnetic analogy to relate ${\bf p}=<{\bf s'}>$ with
$\bf\Omega$ and ${\bf V}$, in the case of simultaneous counterflow
and rotation, but we have not studied the statistic of the
curvature vector $\bf s''$. In contrast, Lipniacki leaves open the
value of $\bf p$ and makes some simple hypotheses about $<|\bf
s''|>$ and $<|\bf s''|^2>$ in his analysis of the possible
influence of polarization in the Vinen's equation.

A further difference between our model and that of Lipniacki
refers to the form of the vortex flux for which he writes
\be%
{\bf J}^L= L{\bf v}^L= L\left[ {\bf v}_s + \alpha {\bf p}\times
{\bf V}+ \beta\alpha I_0 c_{10} (1-|{\bf p}|^2) L^{1/2}{\bf\hat
V}+ \beta\alpha {\bf I_k}L^{1/2} \right]%
\ee where the vector $\bf I_k$ represent the curvature of
 $\vec\omega_s$ lines. This last term is exactly zero if the
 vortex lines are closed, isotropic of straight, and otherwise it
 is expected to be small, except for the case when all the vortex
 lines are parallel to each other but bent. This is only the
 convective contribution, to which it should be added the
 diffusive contribution ${\bf J}^L=-\ti D \nabla L$.

\section{Vortex-line density in steady plane flows}\label{section4}
\setcounter{equation}{0} In equation (\ref{deLsudt-rot})  (and
equation (\ref{deLsudtproposed})) we have rewritten previous
equation (\ref{deLsudt1}) for rotating counterflow turbulence in
liquid helium in terms of $|\textrm{rot} \ {\bf v}_s|$. For pure
rotation, $|\textrm{rot}\ {\bf v}_s|=2 \Omega$ and we just have
our original equation, but (\ref{deLsudt-rot}) may be also used to
describe situations with barycentric motion as plane Couette and
Poiseuille flows (without external heat flux) between two parallel
plates. Here we will consider two plates separated by a distance
$D$, one at rest and the other one moving at velocity ${\bf V}_0$
(Couette flow), or plane Poiseuille flow, given by a longitudinal
pressure gradient along the direction of two parallel quiescent
walls. Here, we will deal with steady states and quasi-stationary
states. We will assume that the flow of the normal component
remains laminar. This requires that the Reynolds number $D
V_0/\eta$, with $\eta$ the viscosity of the normal component and
$V_0$ the characteristic velocity of flow, is sufficiently small.
On the other side, in analogy with the rotating container, we
assume that the velocity $V_0$ is sufficiently high to neglect the
"effects of the walls" \cite{MJ-PRB72-104515}. The essential
problem in both cases is to find the distribution of the
superfluid velocity, the vortex density and the mutual friction
force.  We will focus our attention mainly to steady state
situations, as simple illustration of the changes implied by the
new equations (\ref{deLsudtproposed}) and (\ref{devsdt4}), for $L$
 and ${\bf v}_s$.

\subsection{Plane Couette flow}
We   assume two plane surfaces at $z=0$ and $z=D$ such that the
second one moves parallely to the first one at the velocity ${\bf
V}_0$, and that the relative velocity between normal and
superfluid velocities has a profile ${\bf V}=(V_x(z),0,0)$. The
dynamics of vortex formation is similar to that in the rotating
cylinder. When the upper plate
starts suddenly moving with respect to the lower plate, 
the normal component starts moving under the action of the viscous
force and non-slip condition, whereas the superfluid component
will remain initially insensitive to the motion of the plate.
Thus, a relative velocity (the counterflow velocity) ${\bf V}={\bf
v}_n-{\bf v}_s$ will arise between the two components. This
counterflow velocity ${\bf V}$ depends on the distance from the
lower plate, in particular ${\bf V}$ is maximum for $z=D$ (upper
plate) and zero for $z=0$ (lower plate).

When the counterflow velocity reaches a critical value  near the
moving plane, the remnant vortices, always present in He II, begin
to lengthen and reconnect to form other vortices, which diffuse
towards the lower plate (at rest) forming, in the stationary
situation, an array of vortices parallel to each other and
orthogonal to the flow. Through the vortices, the normal and the
superfluid components become coupled by the mutual friction force
${\bf F}_{ns}$, and the superfluid will tend to match its velocity
with that of the normal fluid ($V\rightarrow 0$); this will
introduce a $\textrm{rot}\ {\bf v}_s\neq 0$ in the superfluid,
expressed by $|\pard {\bf v}_{s}/\pard z|$. After a sufficiently
long time, it is expected that a steady shear flow will have
formed, with ${\bf v}_{n}={\bf v}_{s}$ depending only on $z$ and
having the $x$ direction and such that $\pard {\bf v}_{n}/\pard
z=\pard {\bf v}_{s}/\pard z={\bf V}_0/D$, corresponding to the
physical Newtonian linear profile, which follows from
(\ref{HVBK-sistema-vn}) and (\ref{HVBK-sistema-vs}) with vanishing
tension force $T=0$, and (\ref{F_ns-minus}) in which ${\bf
F}_{ns}={\bf 0}$ for ${\bf V}={\bf 0}$ and $|{\bf p}|=1.$ Then, it
results $|\textrm{rot}\ {\bf v}_s|={\bf V}_0/D$.

Introduction of this value in (\ref{deLsudt-rot}) would give the
areal density of parallel and straight vortex lines, perpendicular
to the flow. However, as it has been said in Section 2, the
replacement of $\Omega$ in terms of $\textrm{rot}\ {\bf v}_s$ is
deeper than a formal substitution because ${\bf v}_s$ will not
become related to the gradient of the barycentric velocity until a
complex transient process has lapsed. Thus, the direct replacement
of $2 \Omega$ in (\ref{deLsudt1}) by $d v_{sx}/dz$ in shear flows,
with $v_{sx}$ the $x$-component of the macroscopic superfluid
velocity, will be valid for steady states and for relatively slow
variations with respect to steady states. Otherwise,
$\textrm{rot}\ {\bf v}_s$ with its own nontrivial dynamics should
be considered in (\ref{deLsudt-rot}). The situation of Couette
flow may be rather illustrative of these features.

Then, the dynamics of $L$ in this case is described by
\be\label{deLsudt-Couette} {dL\over dt}= -\beta \kappa  L^2 +
\left[\alpha_1 V+\beta_2 \sqrt{\frac{\kappa}{2} \left|\frac{\pard
{\bf v}_s}{\pard z}\right|}\right] L^{3/2}-\left[\frac{\beta_1}{2}
\left|\frac{\pard {\bf v}_s}{\pard z}\right|+\beta_4 V \sqrt{
\frac{1}{2\kappa}\left|\frac{\pard {\bf v}_s}{\pard
z}\right|}\right] L-\nabla \cdot {\bf J}^L, \ee where the
coefficients should obey the relations indicated below
(\ref{deLsudt1}), and where the last term stands for the effects
of the vortex flux for inhomogeneous systems.

In the stationary situation $V\approx 0$ and,  according to
(\ref{deLsudt-Couette}), there will be a completely polarized
array of vortices, perpendicular to the velocity and to the
velocity gradient, with uniform areal density given by
\be\label{L=1/k} L=\frac{1}{\kappa}\left|\frac{\pard {\bf
v}_s}{\pard z}\right| = \frac{V_0}{\kappa D}.\ee

 In this view, the stationary character of $L$
would require $V$ to be zero, in such a way that  normal fluid,
superfluid and vortices would move at the same speed and without
internal friction. However, equation (\ref{deLsudt-Couette}) has
the intrinsic feature that for $V$ less than a value $V_{c}$ the
vortex line density does not depend on $V$ and is given by
(\ref{L=1/k}). This critical relative velocity is, according to
(\ref{deLsudt-Couette}),  \be\label{Vc2}
V_{c}=\frac{\beta}{\alpha_1}\left[2\frac{\beta_4}{\alpha_1}-\frac{\beta_2}{\beta}\right]
\sqrt{\frac{\kappa}{2}\left|\frac{\pard {\bf v}_s}{\pard
z}\right|}\cong c' \sqrt{\frac{\kappa}{2}\left|\frac{\pard {\bf
v}_s}{\pard z}\right|},\ee with $c'\approx 3.7$, according to the
values of the coefficients used in (\ref{deLsudt1}) to describe
the value of $V_{c}$ in rotating counterflow velocity.

This indicates that the ordered array of vortices formed in the
Couette flow is stable until $V<V_{c}$. This means that, as $V_0$
grows, the regular array of rectilinear vortices, orthogonal to
$V_0$, is still present and the velocities ${\bf v}_n$, ${\bf
v}_s$ and ${\bf V}$ have  rectilinear profiles, but with slightly
different slope. The value of $V$ is maximum near the plane $z=D$.
When the counterflow velocity $V$ reaches the critical value
$V_{c}$ the regular Couette array of vortices becomes unstable and
a disordered tangle of vortex lines appears between the two plates
in the zone. If one did not apply (\ref{deLsudt1}), but only
intuitive reasoning without the detailed quantitative analysis
showing this critical velocity, one would expect that for $V>0$
will always be a disordered tangle of vortices.

\subsection{Plane Poiseuille flow}
Equation (\ref{deLsudt-Couette}) may be applied to plane
Poiseuille flow between two quiescent parallel walls at $z=\pm
D/2$, driven by a longitudinal pressure gradient. The steady
velocity profile for a Newtonian viscous fluid is parabolic, and
has the form
\be\label{v(x)}%
V_x(z)=\frac{\triangle p}{2 \eta
l}\left[\frac{D^2}{4}-z^2\right]=\frac{\triangle p}{\eta
l}\frac{D^2}{8}\left[1-\frac{4z^2}{D^2}\right]=V_{max}\left[1-\frac{4z^2}{D^2}\right], %
\ee with $\frac{\triangle p}{l}$ the pressure gradient, $\eta$ the
viscosity and $V_{max}=(D^2 \triangle p)/(8 \eta l)$. The pressure
gradient acts on each component in the proportion established by
the HVBK equations
(\ref{HVBK-sistema-vn})--(\ref{HVBK-sistema-vs}).

\begin{figure}
\centering \subfigure[]
{\includegraphics[width=7cm]{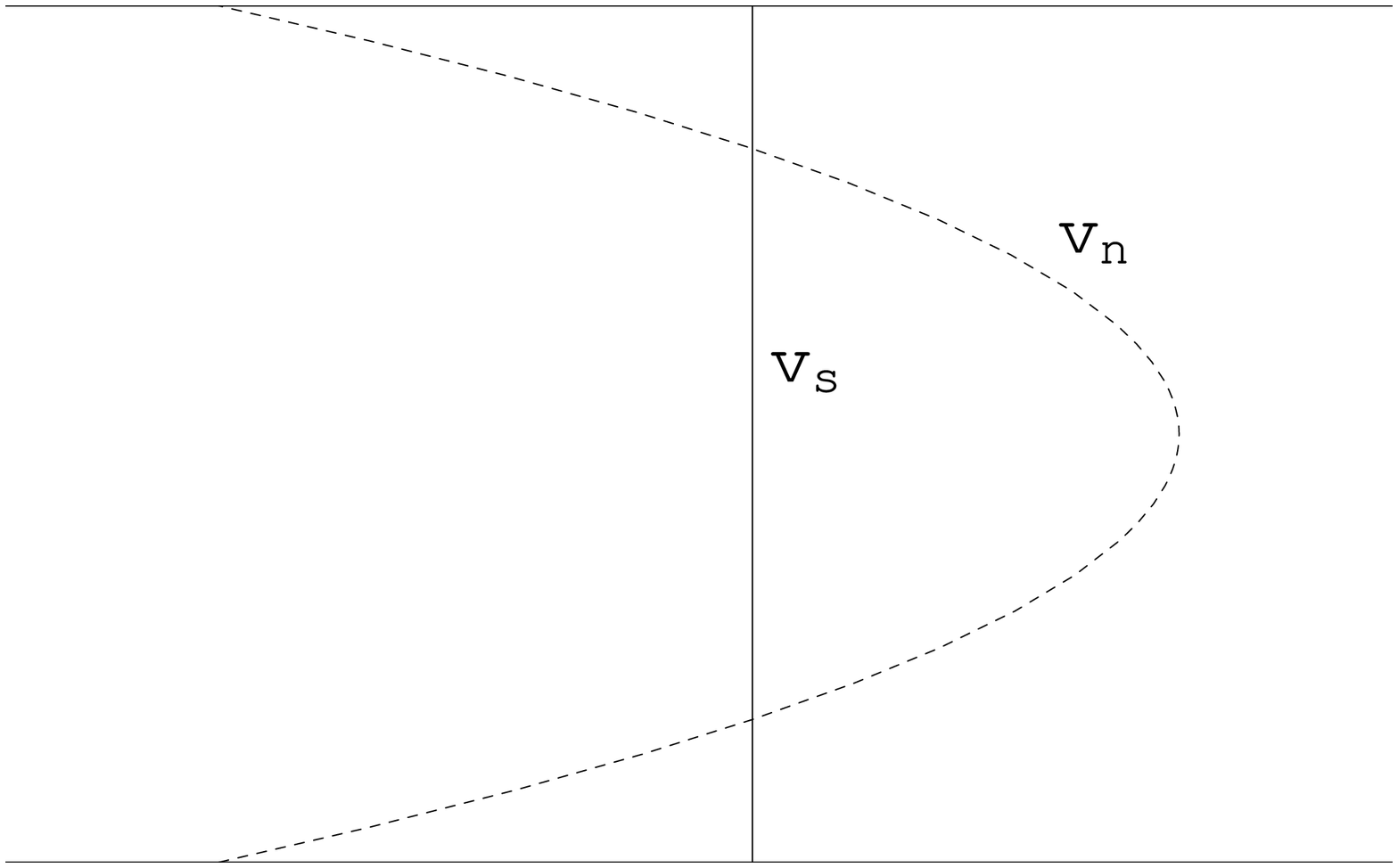}} \hspace{1mm}
\subfigure[] {\includegraphics[width=7cm]{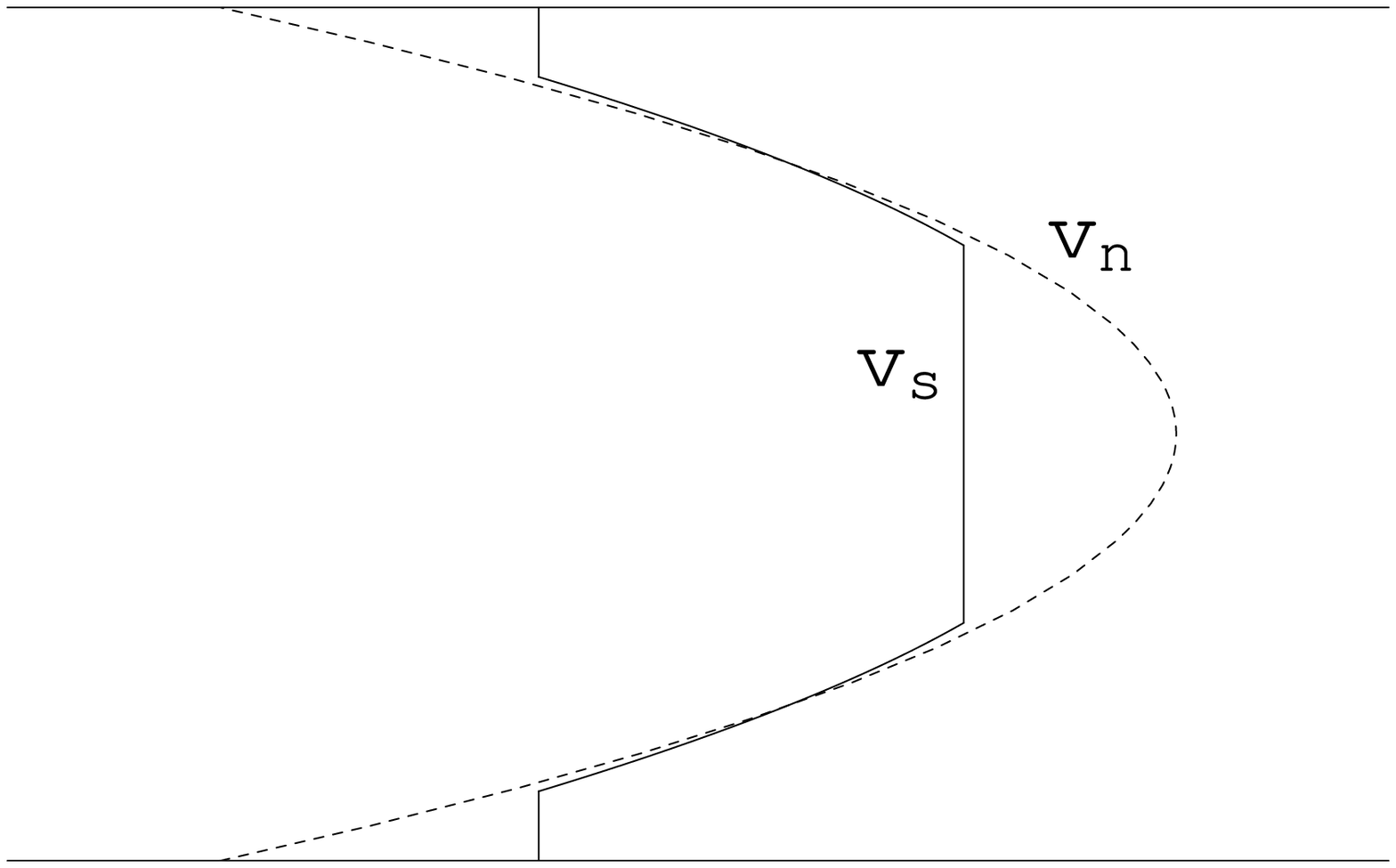}}
\caption{Initial profile (a) and steady profile (b)
  of the superfluid (continuous line) and normal velocities (dashed line).}\label{fig2}
\end{figure}

Initially, the velocity profile of the normal component, submitted
to viscous effects and to no-slip conditions on the walls, will be
rather different from that of the superfluid component, which may
slip freely along the walls (see Figure \ref{fig2}a). As a useful
simplification, one may approximate the velocity profiles as a
parabolic (Poiseuille profile) and a flat profile, respectively
\cite{God_PF13(2001)-983}, which equals one to each other in two
points at the distance $z_0$ from the center of the plates. Then,
one must search how these profiles will evolve under their mutual
interaction due to the friction force, caused by the presence of
the vortices.

During the transient regime, vortices will be produced mainly in
the regions where the relative velocity $V$ is higher than a
critical value $V_c$ --- which may be also the central region ---
but they will be transferred to the matching region where
$v_n=v_s$ because of the second term in expression
(\ref{F_ns-HVBK}) for the  mutual friction force, which is a
Magnus force yielding a vortex lateral drift velocity described by
${\bf v}_{L(lateral drift)}=\alpha' {\bf s}^\prime \times {\bf
V}$. The accumulation of vortices in the region where the two
fluids have the same velocity will enlarge the width of the
matching region (the profile of ${\bf v}_s$ tends to the profile
of ${\bf v}_n$), until arriving at a situation where  $V$ will be
lower than $V_c$ so that not more vortices will be produced. The
steady profile will have the approximate form of Figure
\ref{fig2}b, similar to that considered by Samuels (Fig. 7 of
\cite{Samuel_PRB46(1992)-11714}), but in the matching region ${\bf
v}_n$ and ${\bf v}_s$ are not exactly equal, in contrast with
Couette flow or rotating cylinder, because there is need of a
friction force to cancel out the term in the pressure gradient in
the HVBK equations, as shown in (\ref{HVBK-sistema-vs-P}) below.

In the steady state, for isothermal flow, and neglecting the
tension ${\bf T}$, which vanishes for rectilinear vortices and for
isotropic tangles, equations
(\ref{HVBK-sistema-vn})--(\ref{HVBK-sistema-vs}) reduce to
\bea\label{HVBK-sistema-vn-P}%
 - {\rho_n\over\rho}\nabla
p_n
+ {\bf F}_{ns}+\eta\nabla^2 {\bf v}_n=0,    \\
 - {\rho_s\over\rho}\nabla p_s - {\bf F}_{ns}=0.
\label{HVBK-sistema-vs-P}%
\eea By adding these equations one obtains $- \nabla
p+\eta\nabla^2 {\bf v}_n=0,$ which shows that the velocity profile
of the normal component is the usual one corresponding to the
motion it would have by itself, without the interaction with the
superfluid unless some contributions with ${\bf T}\neq {\bf 0}$
would appear, in the form, for instance, of local anisotropy
vortex tangles. On the other side, from (\ref{HVBK-sistema-vs-P})
it is seen that ${\bf F}_{ns}$ will be different from zero, given
by ${\bf F}_{ns}=-{\rho_s\over\rho}\nabla p_s$. Thus, ${\bf
v}_{n}$ and ${\bf v}_{s}$ will be slightly different, if $\nabla
p$ is low enough, and there will be an array of straight vortices,
which we calculate below.

The most relevant features of the steady profile are: the width $2
z_c$  of the central zone without vortices  and flat ${\bf v}_s$
profile, the width $z_w$ of the boundary layer also without
vortices  and flat ${\bf v}_s$ profile, and $\s L$, the averaged
vortex density in the matching zone where vortices concentrate. We
will compute them from simple qualitative arguments.

To compute $z_c$ and $z_w$ we will ask that the corresponding
circulation of $V_{ns}$ in these regions is lower than the
vorticity quantum $\kappa$. This is a sufficient condition for the
lack of vortices in this zone. The argument is similar to that
which could be used to estimate the critical angular velocity for
the formation of the first vortex line in a rotating cylinder. If
the cylinder is rotating with angular speed $\Omega$, the
circulation of $V$ will be $2 \pi R^2 \Omega$; to obtain
$\Omega_c$ one equates this quantity to $\kappa$, and one gets
$\Omega_c=\kappa /(2\pi R^2).$ The exact result is
$\Omega_c=\kappa \ln (b/a_0)/(2\pi R^2)$ \cite{Do}, with $a_0$ the
radius of the vortex line and $b$ a distance of the order of the
line spacing, which in the case of the one vortex is of the order
of the radius $R$ of the cylinder. In the plane Poiseuille flow
situation the quantity $b$ is of the order $z_c$, in the central
zone, and of the order $z_w$, in the boundary layer zone.

Thus, to estimate $z_c$ we calculate the circulation of
$V_{ns}=\frac{\triangle p}{2 \eta l}\left[z_c^2-z^2\right]$ in the
zone between $z=0$ and $z=z_c$ and equate it to $\kappa \ln
(b/a_0)$. One has
\be\label{zc1}%
\Gamma_c=\oint_{\gamma} V_{ns} \cdot \tr d l=-\int_0^{z_c}
\left.\left(\frac{\triangle p}{2 \eta
l}\left[z_c^2-z^2\right]\right)\right|_{z=0} \tr d
x=\frac{\triangle p}{2\eta
l}z_c^3\approx \kappa \ln (c z_c/a_0)%
\ee where $\gamma$ is the contour of the square whose side is
$z_c$ and $c$ is a numerical constant of the order of the unity.
This may be expressed in terms of the maximum velocity $V_{max}$
of ${\bf v}_n$ as given by (\ref{v(x)}), leading to expression
\be\label{zc2}%
\frac{z_c^3}{D^3}=\frac{\kappa \ln (c z_c/a_0)}{4 D V_{max}}.%
\ee

Concerning the width of the boundary layer $z_w$, a similar
argument yields
\be\label{zw1}%
\Gamma_w=\oint_{\gamma_1} V_{ns} \cdot \tr d l=\int_0^{z_w}
\left.\left(\frac{\triangle p}{2 \eta
l}\left[\left(\frac{D}{2}-z_w\right)^2-z^2\right]\right)\right|_{z=\frac{D}{2}}
\tr d x=\frac{\triangle p}{2\eta
l}\left[Dz_w^2-z_w^3\right]\approx\kappa \ln (c' z_w/a_0),%
\ee where $\gamma_1$ is the contour of the square whose side is
$z_w$ and $c'$ is a numerical constant of the order of the unity.
Up to second order in $z_w$, this yields
\be\label{zw2}%
\frac{\triangle p}{2\eta
l}D z_w^2= \kappa \ln (c' z_w/a_0),%
\ee and using expression (\ref{v(x)}) for the ${\bf v}_n$ profile,
the previous expression may be rewritten in terms of $V_{max}$
 as
\be\label{zw3}%
\frac{z_w^2}{D^2}=\frac{\kappa \ln (c' z_w/a_0)}{4 D V_{max}}.%
\ee This expression  is similar to the one obtained by Samuels in
\cite{Samuel_PRB46(1992)-11714}  for the width of the outer layer
in a cylindrical Poiseuille flow  in a tube of diameter $D$ (his
eq. (15)), which was
\be\label{Samu}%
\left(\frac{r_c}{D}\right)^2=\frac{\kappa }{8\pi D V_{max}}\ln (\frac{8 r_c}{a_0}).%
\ee From (\ref{zc2}) and (\ref{zw3}) it is found that the widths
$z_c$ and $z_w$ decrease for increasing $V_{max}$ as $z_c\sim
V_{max}^{-1/3}$ and $z_w\sim V_{max}^{-1/2}$. Thus for increasing
$V_{max}$ (i.e. increasing pressure gradient) the central zone and
the outer zone boundary layer free of vortices will become
thinner. The flat profile of ${\bf v}_s$ in these zones is
consistent with the absence of vortices, according to the relation
$L=|\pard {\bf v}_s/\pard z|/\kappa$, analogous to the expression
(\ref{L=1/k}), and which vanishes for flat profile.

In the matching region the value of ${\bf v}_n-{\bf v}_s$ will not
be strictly zero, but because of restriction
(\ref{HVBK-sistema-vs-P}) if ${\bf v}_n-{\bf v}_s$ is
approximately constant in this region, one will have that
\be\label{L(z)2}%
L(z)=\frac{1}{\kappa}\left|\frac{\pard {\bf v}_s}{\pard z}\right|\approx
\frac{1}{\kappa}\left|\frac{\pard {\bf v}_n}{\pard z}\right|=\frac{8 V_{max}}{\kappa D^2}|z|.%
 \ee
It is known that there exist two values of $z$ where the
velocities, ${\bf v}_n$ and ${\bf v}_s$, are equal, but, in
general, in the rest of the $z$ domain they could not be exactly
equal. This implies that the mutual friction force could depend on
$z$ and that the distribution of the vortices could not be
homogeneous. To overcome this problem, we average the value of $L$
in the region between $z=z_c$ and $z=z_1=D/2-z_w$. Of course, the
value of $\s L$ in the region between $z=-z_1=-D/2+z_w$ and
$z=-z_c$ will be the same of the first region by symmetry. To
estimate, we assume that the averaged profile of the superfluid
velocity can be approximated by the profile of the normal
velocity, so obtaining
\be\label{L(z)3}%
\s L=\frac{8 V_{max}}{\kappa D^2}\left|\frac{z_c-z_1}{2}\right|=
\frac{2 V_{max}}{\kappa D}\left|1+2\left(\frac{z_c}{D}-\frac{z_w}{D}\right)\right|.%
 \ee
Introducing $z_c$ and $z_w$ as obtained from (\ref{zc2}) and
(\ref{zw3}) we would have an estimate of $\s L$ in terms of
$V_{max}$, or, equivalently, in terms of $\triangle p$. A more
detailed analysis could be carried out from the transversal terms
of the vortex flux, where the Magnus drift and the diffusion flux
in (\ref{J^L}) would cancel each other.

Expression (\ref{zc2}) may be used to obtain the conditions for a
laminar flow without any vortex. This situation will be found when
the width of the central zone without vortices $z_c$ is bigger
than $D/2$. This leads to the condition $D V_{max}/\kappa\leq 2
\ln \left(D/(2 a_0)\right)$. For $D\approx 10^{-2}\textrm{m},$ and
since $a_0\approx 10^{-10}\textrm{m}$, we have $D
V_{max}/\kappa\leq 20.$ The dimensionless quantity $D
V_{max}/\kappa$ is analogous to the Reynolds number. In viscous
fluid, the Reynolds number is defined as $DV/\nu$, with $\nu$
being the kinematic viscosity $\nu=\eta /\rho$, which has
dimensions $\textrm{m}^2\textrm{s}^{-1}$. The vorticity quantum
$\kappa$ has also dimension $\textrm{m}^2\textrm{s}^{-1}$ and
therefore  $D V_{max}/\kappa$ may be considered as a quantum
Reynolds number. A similar number has been used in pure
counterflow experiments in cylindrical containers of diameter $D$
where, for instance, the appearance of the first vortex takes
place at $T=1.7 \textrm{K}$ for $D V/\kappa \approx 80$
\cite{MJ-JPCM17-2005}. A more rigorous stability analysis would be
convenient to obtain more values of the critical quantum Reynolds
number in both situations.

\section{Conclusions}
The quantized character of vorticity in superfluids emphasizes the
special importance of vortex lines, whose dynamics becomes a
central aspect of rotating or turbulent flows of superfluids. The
main proposal of this paper is equation (\ref{deLsudt-rot}) for
the evolution of vortex line density, which generalizes our
previous proposal (\ref{deLsudt1}) for rotating counterflow
situations. Here, by writing the local average rotational of the
superfluid component instead of the angular velocity, we have
enlarged the set of applications of the theory in two main
aspects. One of them is that (\ref{deLsudt-rot}), in contrast to
(\ref{deLsudt1}), may be applied not only to rotation but also to
shear flows, as illustrated in Section~\ref{section4}. The second
enlargement is of dynamical nature: in (\ref{deLsudt-rot})
$\textrm{rot}\ {\bf v}_s$ appears, and ${\bf v}_s$ itself should
satisfy its own evolution equation, which is coupled to the
evolution of  ${\bf v}_n$, the velocity of the normal component.
Then, (\ref{deLsudt-rot}) becomes deeply coupled to the HVBK
equations (\ref{HVBK-sistema-vn}) and (\ref{HVBK-sistema-vs}) for
${\bf v}_n$ and  ${\bf v}_s$ not only through the mutual force
${\bf F}_{ns}$ between the normal and the superfluid components,
which requires the knowledge of $L$, but also because in
(\ref{deLsudt-rot})  ${\bf v}_s$ is needed to obtain $L$. Thus,
the coupling of these equations is much emphasized in
(\ref{deLsudt-rot}) as compared to (\ref{deLsudt1}).

For situations close to nonequilibrium steady states or for slow
variations of ${\bf v}_s$, in such a way that $\textrm{rot}\ {\bf
v}_s$  is well described by the angular velocity or by the
barycentric velocity gradient, equations (\ref{deLsudt1}) or
(\ref{deLsudt-Couette}) describe the vortex line density in terms
of $\Omega$ or $d v_{sx}/dz$. In each case we have provided an
estimation of the vortex density and of the superfluid velocity
profile in the steady state.

We have compared our proposal with that of Lipniacki, which shares
the objectives of the present paper but stresses the polarization
${\bf p} = \textrm{rot}\ {\bf v}_s/kL$ more than $\textrm{rot}\
{\bf v}_s$ itself. Lipniacki's evolution equation for $L$ is,
essentially, the classical Vinen's equation, but with the new
aspect that its coefficients become dependent on the polarization
$\bf p$ according to the microscopic identification of the
coefficients proposed by Schwarz \cite{SchwPRB38-1988}. The
disagreement between our equation (\ref{deLsudtproposed}) and the
Lipniacki's proposal (\ref{deLsudtLipni}) may be due to the
different physical origin of the terms dependent on the
polarization. Our opinion is that Schwarz derivation of Vinen's
equation (\ref{deLsudtLipni2}) does not include some relevant
contributions of rotational systems. For straight parallel
vortices, as those arising in pure rotation experiments, both the
production and the destruction terms vanish. This is consistent
with Schwarz's postulates for the vortices, but in purely
rotational flows the dynamics of vortices has a different origin,
related to the migration of vortices formed on the wall towards
the center of the system, and with repulsion forces amongst
vortices. Thus a general treatment would require to include these
effects besides the Scharwz effects in (\ref{deLsudtLipni2}), and
it could provide a further understanding of the differences
between (\ref{deLsudtproposed}) and (\ref{deLsudtLipni}). In any
case, comparison with experimental results in Fig. 1 indicates
that (\ref{deLsudtproposed}) yields a better description of them.

\section*{Acknowledgments}
We acknowledge the support of the Acci\'{o}n Integrada
Espa\~{n}a-Italia (Grant S2800082F HI2004-0316 of the Spanish
Ministry of Science and Technology and grant IT2253 of the Italian
MIUR). DJ acknowledges the financial support from the
Direcci\'{o}n General de Investigaci\'{o}n of the Spanish Ministry
of Education under grant FIS 2006-12296-C02-01 and of the
Direcci\'{o} General de Recerca of the Generalitat of Catalonia,
under grant 2005 SGR-00087. MSM and MS acknowledge the financial
support from MIUR under  grant "PRIN 2005 17439-003" and by "Fondi
60\%" of the University of Palermo. MS acknowledges the "Assegno
di ricerca: Studio della turbolenza superfluida e della sua
evoluzione" of the University of Palermo.

\end{document}